\documentclass[aps,tightenlines,showpacs]{revtex4}
\usepackage{graphicx}
\usepackage{float}
\usepackage{amssymb}

\newcommand{\bea}{\begin{eqnarray}}
\newcommand{\eea}{\end{eqnarray}}
\newcommand{\xx}{\noindent}

\newcommand{\Psl}{P\!\!\!\!/~}

\newcommand{\Rbarsl}{\bar R\!\!\!\!/~}
\begin{document}

\title{\Large{Energetic di-leptons from the Quark Gluon Plasma}}

\author{M.E. Carrington and A. Gynther}

\affiliation
{Department of Physics, Brandon University,
Brandon, Manitoba, R7A 6A9 Canada\\
and Winnipeg Institute for Theoretical Physics,
Winnipeg, Manitoba, Canada}

\author{P. Aurenche}

\affiliation
{LAPTH, Universit\'{e} de Savoie, CNRS, 9 Chemin de Bellevue, B.P. 110, F-74941,France}

{\it \today}

\begin{abstract}
In this paper we study the production of energetic di-leptons.  
We calculate the rate for 2 $\to$ 2 processes. The log term is obtained analytically and the constant term is calculated numerically. When the photon mass is of the order of the thermal quark mass, the result is insensitive to the photon mass and the soft logarithmic divergence is regulated by the thermal quark mass, exactly as in the case of real photons. 
We also consider the production
of thermal Drell-Yan dileptons (thermal quark and antiquark pairs produced by virtual photons) and calculate the rate systematically in the context of the hard thermal loop effective theory.  We obtain analytic and numerical results. We compare our results with those of previous calculations.

\end{abstract}

\pacs{11.10.Wx, 12.38.Mh, 25.75.Cj}
\maketitle
\section{Introduction}

It is believed that experiments at RHIC are producing quark-gluon plasma. The properties of the plasma will be further studied by the heavy ion program at the LHC. Thermalization and rescattering tend to erase information about the state of the plasma at early times, and therefore it is of interest to study high energy photons which are produced early in the collision and escape without further interaction (for a review, see \cite{CERN}).

The main disadvantage of using photons as a signal is that there are several sources of photons, and it is necessary to separate the signal from the background.
Photons that are produced in the early stages of the collision are called direct photons. There are two types of direct photons: prompt photons, which are produced in the initial collisions of the partons which make  up the heavy ions, and thermal photons, which are produced in the hot quark or hadronic matter formed during the collision.  
Prompt photons have a power damped spectrum and dominate over thermal photons at large energies.
Thermal photons produce an exponentially damped spectrum and dominate prompt photons at lower energies. 
In addition to direct photons, there is a large background contribution from photons produced through the radiative decay of hadrons.  These photons do not provide direct information on the early stages of the quark-gluon plasma.

Of these three kinds of photons (prompt/thermal/decay), we are primarily
interested in thermal photons, since they are the ones that provide direct
information on the early stages of the plasma. In order to isolate the
contribution of thermal photons, one should look at not too large transverse momentum, typically a few GeV's. To overcome the background problem 
in this range, it is of interest to study the spectrum of small mass 
di-leptons (virtual photons) which have the same production mechanisms as real photons.  In fact, the PHENIX experiment at RHIC has been able to extend the accessible $q_T$ range of the photon spectrum down to 1 GeV using this approach~\cite{phenix}. Throughout this paper we use $q_0$ for
the photon energy, $Q:=\sqrt{Q^2}~~(Q^2=q_0^2-q^2)$ for the invariant mass, and $T$ for temperature. We focus on the case $q_0\gg T\gg Q$. 

In general, the production rate for real or virtual photons is obtained from the
imaginary part of the contracted retarded photon polarization tensor (see Eqns.
(\ref{rate1}) and (\ref{rate2})). We work in the
framework of the ``Hard Thermal Loop'' (HTL) resummed perturbation
theory~\cite{BraatenPis} where the quarks and gluons acquire an effective mass
as well as a dispersive component.

For real photons, there are two physically distinct contributions to the
polarization tensor at leading order. 

Real photons are produced through Compton scattering and 2 $\to$ 2 annihilation
processes. Collectively, these will be called 2 $\to$ 2 processes. They are
obtained from the central, or real, cuts of the two loop diagrams, where the
gluon is time-like and on shell. The 2-loop contributions to the polarization
tensor are shown in Fig. \ref{caFig}. The amplitudes that are obtained from the
real cuts with time-like gluons are shown in Fig. {\ref{caAmp}. 
\par\begin{figure}[H]
\begin{center}
\includegraphics[width=7cm]{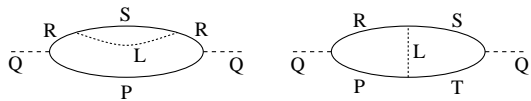}
\end{center}
\caption{2-loop contributions to the polarization tensor: solid lines are quarks, dashed lines are photons and dotted lines are gluons.}
\label{caFig}
\end{figure}
\par\begin{figure}[H]
\begin{center}
\includegraphics[width=6cm]{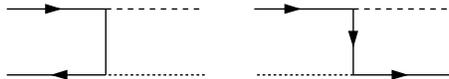}
\end{center}
\caption{Amplitudes that correspond to 2 $\to$ 2 processes. Crossed diagrams are
not shown. }
\label{caAmp}
\end{figure}
In addition, there is a leading order contribution to the rate from
bremsstrahlung and off-shell annihilation processes. These processes contribute
at leading order because of a strong collinear enhancement. They are obtained
from the real cuts of the two loop diagrams in Fig. \ref{caFig}, where the gluon is taken to be HTL resummed and space-like. The scattering amplitudes are shown in Fig. \ref{bremFig}.
\par\begin{figure}[H]
\begin{center}
\includegraphics[width=5cm]{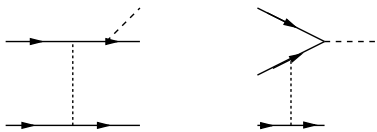}
\end{center}
\caption{Amplitudes for bremsstrahlung and off-shell annihilation.}
\label{bremFig}
\end{figure}
For the case of space-like gluons, higher order diagrams involve the same
enhancement and therefore also contribute at leading order. This is known as the
LPM effect. The diagrams are resummed using a set of integral equations \cite{AMY}. Real
cuts with space-like gluons produce a set of amplitudes that corresponding to
multiple rescatterings.  For off-shell annihilation processes, some of these amplitudes
are shown in Fig. \ref{lpmFig}. The rate obtained from Eqn. (\ref{rate2}) involves
terms that come from squaring these amplitudes, and also terms that correspond
to interference effects. 
\par\begin{figure}[H]
\begin{center}
\includegraphics[width=7cm]{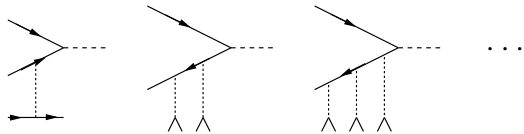}
\end{center}
\caption{Some multiple rescattering diagrams.}
\label{lpmFig}
\end{figure}

Now we discuss the case of virtual photons. In this case a new scale $Q$ enters
the calculation. The complete calculation has already been done, but due to
several misunderstandings, the pieces have not been correctly assembled. 

We start with the observation that in HTL the quark acquires an effective mass $m_\infty$, and the imaginary part of the 1-loop diagram has
a threshold at $Q^2>4m_\infty^2$  for a thermal Drell-Yan processes of the form $q\bar q \to \gamma$, which corresponds to a thermal quark anti-quark pair produced from a virtual photon. The self energy diagram and the amplitude that is
extracted from the imaginary part are shown in Fig. \ref{dyFig}. \footnote{The
threshold for processes of the form $q \to q+\gamma$ is $Q^2<(m_1-m_2)^2$.
As is discussed in section \ref{DY}, both quarks carry HTL propagators where the
momentum is taken to be much larger than the thermal quark mass. In this limit,
$\omega_+ \approx \sqrt{p^2+m_\infty^2}$ and $\omega_-\approx p$ and therefore
the process $q \to q+ \gamma$ must involve one plus mode and one minus mode.
Since the residue of the minus mode is exponentially suppressed at large
momentum, there is no phase space for this process.}.
\par\begin{figure}[H]
\begin{center}
\includegraphics[width=7cm]{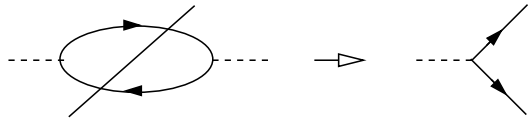}
\end{center}
\caption{The lowest order thermal Drell-Yan process.}
\label{dyFig}
\end{figure}
There will also be contributions from the side cuts of the 2-loop diagrams shown in Fig. \ref{caFig}, and from the side cuts of all contributions to the polarization tensor that produce the amplitudes shown in Fig. \ref{lpmFig}. 
When the gluons are taken to be space-like, these side cuts give contributions that correspond to interference between the tree level amplitude and virtual corrections to it.  Some of the amplitudes are shown in Fig. \ref{oddFig} (again, we show only annihilation processes). The dilepton rate obtained from Eqn. (\ref{rate1}) involves terms that come from squaring these amplitudes, and also terms that correspond to interference effects.
\par\begin{figure}[H]
\begin{center}
\includegraphics[width=7cm]{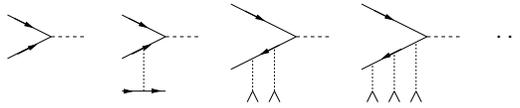}
\end{center}
\caption{The amplitudes involved in the LPM resummation for virtual photons}
\label{oddFig}
\end{figure}
In principle, there is also a contribution at leading order from the side cut of
the 2-loop diagrams where the gluon is time-like. The contribution from soft
time-like gluons is subleading. The contribution from hard time-like gluons is
included by using asymptotic propagators and vertices in the self energy
diagrams that produce the amplitudes shown in Fig. \ref{oddFig}. This point will
be discussed in detail in section \ref{DY}. \\

The results for these contributions that have been obtained by previous
calculations are summarized below.

For real photons, the rate for 2 $\to$ 2  processes has been calculated in Refs.
\cite{joe,rolf}.  

For real photons the rate for bremsstrahlung and off-shell annihilation
processes  has been calculated in Ref. \cite{AMY}. 

For virtual photons the rate for bremsstrahlung and off-shell annihilation
processes has been calculated in Ref.
\cite{PAbrem2}. The leading term in the LPM resummation is the rate for the
thermal Drell-Yan $q\bar q\to\gamma$ process.

In Ref.
\cite{ruus} the thermal Drell-Yan contribution was calculated to logarithmic accuracy assuming $q_0\gg Q\gg m_f$ (where $m_f$ is the quark thermal mass).  A comparison with our calculation is given in section \ref{ruus-c}.

The thermal Drell-Yan contribution was also calculated in Ref. \cite{TT}. These authors attempt to
produce a result that is valid at next-to-leading order for $q_0\gg T\gg \{m_f, Q\}$.  There are several
problems with this calculation. A detailed discussion is given in section \ref{TT-c}. \\

This paper is organized as follows: In section \ref{notation} we define our notation. 

In section \ref{ca} we calculate the rate for 2 $\to$ 2  processes for virtual photons.
We
include the new scale $Q$, and an asymptotic mass for the hard fermion. We show
that the asymptotic mass does not affect the result. The log term is obtained analytically and the constant term is calculated numerically. Our result correctly reduces to the result for real photons, in the limit that the photon mass goes to zero. 

In section \ref{DY} we calculate the contribution to the rate from thermal Drell-Yan processes both
analytically and numerically. The numerical result is valid for $q_0\gg T\gg
\{m_f,Q\}$, without any restrictions on the relative size of $Q$ and $m_f$. We produce two analytic expressions: one is valid for small $m_f$ and one is valid close to the threshold.

In section \ref{COMP} we compare our results with those of previous calculations. 

In section \ref{conc} we present our conclusions.

\section{Notation}
\label{notation}

In this section we define our notation. We write $\text{Im}\Pi:=\text{Im}\,(\Pi_{{\rm ret}})^\mu_{~~\mu}$. The rate for virtual photons (di-leptons) is:
\bea
\label{rate1}
{\rm rate}:=\frac{d^4\Gamma}{d^4x\, d^4q}=\frac{e^{-\beta  q_0}\alpha }{12\pi ^4 Q^2}\text{Im}\Pi.
\eea
The rate for real photons is given by a similar expression: 
\bea
\label{rate2}
{\rm rate}:=q_0\cdot \frac{d^3\Gamma}{d^4 x\,d^3q} =\frac{e^{-\beta  q_0}} {(2\pi)^3}\text{Im}\Pi.
\eea

We use conventional notation for the thermal distribution functions:
\bea
&& n_b(p_0)=\frac{1}{e^{p_0\beta}-1}\,;~~~~n_f(p_0)=\frac{1}{e^{p_0\beta}+1}. 
\eea
We write 4-vectors as capital letters, for example $P=(p_0,\vec p)$. In addition, we give specific names to particular combinations of momenta (see Fig. \ref{caFig}):
\bea
S= L+P+Q,~~T= L+P,~~R= P+Q,~~S^\prime=-S. \nonumber
\eea
We define angles as follows:
\bea
&&\vec q=(0,0,q),\\
&&\vec p=(p\sin\beta,0,p\cos\beta),\nonumber\\
&&s^\prime=(s^\prime \cos\phi\sin\gamma,s^\prime \sin\phi\sin\gamma,s^\prime \cos\gamma).\nonumber
\eea
We write the cosines of these angles as:
\bea
\cos\beta=x_\beta\,,~~\cos\gamma=x_\gamma.
\eea
For bare on shell fermions and gauge bosons (in the Feynman gauge) we write:
\bea
\label{bareProp}
&&S_{\rm ret}(P)-S_{\rm adv}(P)= \Psl d(P), \\[2mm]
&&D_{\rm ret}^{\mu\nu}(L)-D_{\rm adv}^{\mu\nu}(L)=-g^{\mu\nu}d(L),\nonumber\\[2mm]
&&d(X)=-2\pi i \,{\rm Sign}(x_0)\;\delta(X^2).\nonumber
\nonumber
\eea
For a HTL fermion:
\bea
S_{\rm ret}(P)&&=\frac{1}{2}\left((\gamma_0-\hat p^i \gamma^i)\Delta_{\rm ret}^+(P)+(\gamma_0+\hat p^i \gamma^i)\Delta_{\rm ret}^-(P)\right).
\eea
The spectral density is obtained from:
\bea
\label{htl}
\rho_\pm(P) &&= i d_\pm (P)=i(\Delta_{\rm ret}^+(P) - \Delta_{\rm adv}^+(P)), \\[2mm]
d(P) &&=-2\pi i \sum_{j=\pm 1} Z_\pm\delta(p_0-j \omega_\pm)-2 \pi i \beta(p_0,p), \nonumber\\[2mm]
Z_\pm&&=\frac{\omega_\pm^2-p^2}{2m_f^2}\,,~~~\beta_+(p_0,p) = \beta_-(-p_0,p)\,,~~~m_f^2=\frac{1}{2}C_F \,\alpha_s\,\pi \,T^2\nonumber\\
\beta_+(p_0,p)&&=\frac{2 p^2 m_f^2 \left(p-p_0\right) \Theta \left(p^2-p_0^2\right)}{\pi ^2 \left(p-p_0\right)^2
   m_f^4+\left(2 \left(p-p_0\right) p^2+m_f^2 \left(p
\left(\text{ln}\left|\frac{p+p_0}{p_0-p}\right|+2\right)-\text{ln}\left|\frac{p+p_0}{p_0-p}\right| p_0\right)\right)^2}.\nonumber
\eea

\section{Compton and Annihilation Processes}
\label{ca}

We expect that leading order contributions to Compton and annihilation processes come from the real, or central, cuts of the diagrams in Fig. \ref{caFig}, where the gluons are time-like. 
The first diagram in this figure is referred to as the self-energy diagram and the second is called the vertex diagram. We note that there are two self-energy diagrams, which correspond to correcting either the top or bottom fermion line. Both diagrams give the same contribution and thus we could use either one of them with a factor of two but, in order to obtain a more symmetric expression, we explicitly include both diagrams. 

Only the hard part of the gluon momentum contributes, and thus it is safe to use a bare on-shell gluon propagator. On the other hand, both hard and soft quark momenta contribute. In order to calculate the integral, we separate the phase space into two regions by introducing a scale $p_c \sim \sqrt{g}T$. We calculate separately the contributions involving exchanged quarks with momenta greater than and less than $p_c$. Different approximations are used in each region. The separation scale $p_c$ cancels when the two pieces are combined. This technique was introduced in \cite{Braa}.

At zero temperature, the integral corresponding to the diagrams in Fig. \ref{caFig} can be written:
\bea
{\rm Im}\Pi&&=-\frac{e^2\,g^2\,  C_F N_c \,e_q^2}{(2\pi)^8}\int dP\,\int dS\\
&&\cdot\;\Big(2{\rm Tr}\big({\cal S}(P)\gamma_\mu {\cal S}(R)\gamma_\tau {\cal S}(S) \gamma_\lambda {\cal S}(R) \gamma^\mu\big)+{\rm Tr}\big({\cal S}(P)\gamma_\tau {\cal S}(T)\gamma_\mu {\cal S}(S)\gamma_\lambda {\cal S}(R)\gamma^\mu\big)\Big)D^{\tau\lambda}(L),\nonumber
\eea
where $e_q^2=5/9$, $C_F=4/3$, $N_c=3$, $\int dP:=\int dp_0\int d^3p$, and $i{\cal S}$ and $-iD_{\mu\nu}$ correspond to the quark and gluon lines respectively. We need to obtain the corresponding integral at finite temperature. We work in the Keldysh representation of the real time formalism. The method we use to sum over Keldysh indices is described in \cite{MCTF} and our technique for taking the imaginary part is found in \cite{Hou1}. We calculate the contribution for hard exchanged quarks by explicitly imposing a lower limit cutoff on the quark momentum. The calculation is completely insensitive to soft momentum scales, which means that we can use bare propagators (Eqn. (\ref{bareProp})).  In addition, we drop terms in the numerator that are proportional to $Q^2$, since we have assumed $q_0\gg\{Q,m_f\}$ throughout. The result for the real cuts is:
\bea
\label{2L-int}
{\rm Im}\Pi&&=\frac{\alpha \,\alpha _s \, C_F N_c \,e_q^2}{2 \pi ^5}\int dp\,p^2\int dp_0 \int ds^\prime\;(s^\prime)^2 \int ds^\prime_0 \int dx_\beta \int dx_\gamma \int d\phi ~~\big[\text{SE}+\text{VER}\big]\\[4mm]
&&\cdot \;d_L d_{S^\prime} d_P \left(\left(1+n_b\left(l_0\right)\right) \left(1-n_f(s_0^\prime)\right)
   \left(1-n_f\left(p_0\right)\right)-n_b\left(l_0\right) n_f(s_0^\prime) n_f\left(p_0\right) \right),\nonumber\\[4mm]
&&\text{SE} = \frac{i \left(R^4+4 T^2 R^2+T^4\right)}{R^2 T^2} \,;~~~~ \text{VER} = -\frac{2 i \left(R^2+T^2\right)^2}{R^2 T^2}\,. \nonumber
\eea
The term $\text{SE}$ corresponds to the first diagram in Fig. \ref{caFig}, combined with the diagram with the bottom fermion corrected. The term $\text{VER}$ corresponds to the second diagram in Fig. \ref{caFig}. 
  The result is exactly the same for real and virtual photons. Following \cite{rolf} we obtain:
\bea
\label{RH}
{\rm Im}\Pi_{2\to 2}^{\rm hard}=2 \alpha  N_c e_q^2 m_f^2 \left(\ln \left(\frac{T q_0}{p_c^2}\right)
+C^{\rm hard}_{2\to 2}\right)\,;~~~~C^{\rm hard}_{2\to 2}=\frac{\zeta^\prime(2)}{\zeta(2)}+\frac{\ln
   (2)}{3}-\gamma +\frac{3}{2}\,.
\eea

For soft exchanged quarks a slightly different technique is necessary 
since the integral diverges when the virtuality of the exchanged quark ($R^2$ or
$T^2$) goes to zero.  This is just the expected result that the cross section
involving the exchange of a massless particle is infinite.  The well known
solution to this problem is to replace the soft quark propagator with the
corresponding HTL propagator. Effectively, when the exchanged quark is soft, the
diagrams in Fig. \ref{caFig} are replaced with the diagrams  shown in Fig.
\ref{1-loop}, where the propagator with the solid dot is the HTL quark
propagator and the bare line is a hard quark propagator. Using symmetry, we
calculate the first diagram and multiply by a factor of two.
\par\begin{figure}[H]
\begin{center}
\includegraphics[width=6cm]{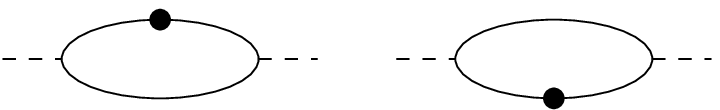}
\end{center}
\caption{1-loop contributions to Compton and annihilation processes. The quark lines carry momenta $P$ and $R=P+Q$.}
\label{1-loop}
\end{figure}

The integral corresponding to two times the first diagram in Fig. \ref{1-loop} is given in Eqn. (\ref{1L-int}). The dressed line carries momentum $P$ and the hard line has momentum $R=P+Q$. 
\bea
\label{1L-int}
{\rm Im}\Pi&&=~\frac{4 i \pi  \alpha   N_c e_q^2}{\pi^2} ~\int dp_0\int dp\;p^4\int dx_\beta~\left(n_f\left(p_0\right)-n_f\left(p_0+q_0\right)\right)\;\delta \left(\left(p_0+q_0\right)^2-\omega(r)^2\right)\\[2mm]
&&~~~~ \left(d_+\left(p_0,p\right) \left(-p+p_0+q_0-q x_{\beta
   }\right)+d_-\left(p_0,p\right) \left(p+p_0+q_0+q x_{\beta }\right)\right) \nonumber
\eea
The real cut is obtained from the cut part of the HTL dispersion relation $d_\pm(p_0,p) \to -2 \pi i \beta_\pm(p_0,p)$.  Since there is an explicit factor $m_f^2\sim \alpha_s T^2$ in the numerators of the factors $\beta_\pm(p_0,p)$ (see Eqn. (\ref{htl})), we will drop all additional factors of order $\{m_f,Q\}$ in the numerator of (\ref{1L-int}).

We use the delta function to do the angular integral. Setting the argument of
the delta function to zero and solving for $x_\beta$ gives the solution $\tilde
x_\beta$. Writing $p_0 = z p$ we define $g(z,p):=1-\tilde x_\beta^2$ and obtain
the phase space constraint $g(z,p)>0$, which we can write as a theta function.
The limits on the $p$-integral are determined from this constraint. We use $\omega(r)=\sqrt{r^2+m_\infty^2}=\sqrt{p^2+2 q \,p\,x_\beta+q^2+m_\infty^2}$ where we define the asymptotic quark mass by $m^2_\infty=2m_f^2$. 
Doing the angular integral we obtain:
\bea
\label{RS-full}
{\rm Im}\Pi_{2\to 2}^{\rm soft} &&=
8\,\alpha\,\,N_c\,e_q^2\;m_f^2\,  \int_0^{p_c} \int ^1_{-1}dz dp\;\left(n_f(p z)-n_f\left(p z+q_0\right)\right)\,\big(f(z,p)+f(-z,p)\big)\,\Theta\big(g(p,z)\big),\\[2mm]
f(z,p)&&= \frac{p^3 (z+1)^2 }{\pi ^2 (z + 1)^2 m_f^4+\left( (z+1) \ln
   \left(\frac{z+1}{1-z}\right) m_f^2-2 \left((z+1) p^2+m_f^2\right)\right)^2}, \nonumber
\eea
where we have defined:
\bea
\label{theta}
&&g(z,p) = 4 p^2 q^2-\left(\left(z^2-1\right) p^2+2 z q_0 p+Q^2-  m_\infty^2\right)^2\,.
\eea

Since $T\gg p_c>p$ we can use
\bea
n_f(pz)-n_f(p z+q_0)~ \approx~ \frac{1}{2}-n_f(q_0))~\approx ~\frac{1}{2}
\eea
to factor the thermal functions out of the integrand and obtain:
\bea
\label{piTemp}
{\rm Im}\Pi_{2\to 2}^{\rm soft} &&=
4\,\alpha\,\,N_c\,e_q^2\;m_f^2\,  \int_0^{p_c} \int ^1_{-1}dz dp\;\big(f(z,p)+f(-z,p)\big)\,\Theta\big(g(p,z)\big)\,.
\eea
Corrections to this result are or order $p_c/T$ and are thus of the same order as terms that have been dropped in the calculation of the contribution from hard exchanged quarks. Recall that we are working in the limit $q_0\gg T\gg p_c \gg \{m_f,Q\}$, and it is only in this limit that the dependence on the cutoff cancels between the contributions from hard and soft exchanged quarks. 

Using $q_0 \gg \{m_f,Q\}$ we can rewrite the constraint given by the theta function in terms of integration limits:
\bea
\begin{array}{lll}
&   & - 1 < z < 1 - {Q^2 - m_\infty^2 \over 2 q_0 p} ~~{\mbox {\rm for}} ~~ Q^2 - m_\infty^2 > 0 \\
{|Q^2 - m_\infty^2| \over 4 q_0}  < p < p_c &  ~~~~~~~~{\rm and}~~~~~~~~  &  \\
& &  - 1 + {m_\infty^2 - Q^2 \over 2 q_0 p} < z < 1 ~~{\mbox {\rm for}} ~~m_\infty^2 - Q^2> 0  \\
\end{array}
\eea
Using these results, for $m_f/q_0\ll 1$, the integral is completely parametrized by the dimensionless variable $\xi:\,=Q^2/(q_0m_f)$.

The logarithmic term is extracted by taking the static limit $z \to
0$. The plus and minus modes give the same contribution to the log term. The result is:
\bea
\label{RS-log}
{\rm Im}\Pi_{2\to 2}^{\rm soft}= 2\,\alpha\,N_c\,e_q^2\;m_f^2\, \left[\frac{1}{2}\ln \left(\frac{p_c^4}{\frac{\pi^2}{4}m_f^4
+\big[m_f^2+\frac{(Q^2-m_\infty^2)^2}{4q_0^2}\big]^2}\right) + C^{\rm soft}_{2\to 2}(\xi)\right]
\eea
where $C^{\rm soft}_{2\to 2}(\xi)$ is a constant that must be determined numerically. 
Combining (\ref{RH}) and (\ref{RS-log}) we see that the cutoff $p_c$ cancels. We obtain:
\bea
\label{REALx}
{\rm Im}\Pi_{2\to 2} = 2\,\alpha\,N_c\,e_q^2\;m_f^2\, \left[\frac{1}{2}\ln \left(\frac{T^2 q_0^2}{\frac{\pi^2}{4}m_f^4
+\big[m_f^2+\frac{(Q^2-m_\infty^2)^2}{4q_0^2}\big]^2}\right) +C^{\rm hard}_{2\to 2} +C^{\rm soft}_{2\to 2}(\xi)\right]\,.
\eea
We consider the denominator of the log in Eqn. (\ref{REALx}). For $q_0\gg Q\sim m_f$ $(\xi \ll 1)$ we can drop the second term in the square brackets and the expression reduces to the result of \cite{joe,rolf} for real photons. However, we must be careful when $\xi$ is large.  It is easy to see that the terms that are proportional to the asymptotic mass are always negligable (by powers of $Q/q_0$) compared to terms proportional to the thermal mass. This result indicates that it is not necessary to include the asymptotic mass on the hard line in Fig. \ref{1-loop}, or equivalently, that the collinear part of the integral is not dominant. 
Dropping the asymptotic mass we obtain:
\bea
\label{REAL}
{\rm Im}\Pi_{2\to 2} = 2\,\alpha\,N_c\,e_q^2\;m_f^2\, \left[\frac{1}{2} \ln \left(\frac{T^2 q_0^2}{\frac{\pi^2}{4}m_f^4
+\big[m_f^2+\frac{Q^4}{4q_0^2}\big]^2}\right) +C^{\rm hard}_{2\to 2} +C^{\rm soft}_{2\to 2}(\xi)\right]\,.
\eea
Note that for $\xi \sim 1$  all three terms in the denominator of the log are of the same order. \\

For $\xi=0$ the numerical result for the constant $C^{\rm soft}_{2\to 2}(\xi)$ agrees with the result for real photons \cite{joe,rolf}. 

For $\xi \gg 1$ we can obtain an analytic expression (up to corrections of order $Q^2/q_0p_c$) for the integral in Eqn. (\ref{piTemp}):
\bea
\label{tempY}
{\rm Im}\Pi_{2\to 2}^{\rm soft} &&=
2\,\alpha\,N_c\,e_q^2 \, m_f^2\,  \int_{Q^2/(4q_0)}^{p_c}\frac{dp}{p} \int ^{1-Q^2/(2p q_0)}_{-1}dz \\
&& = 
2\,\alpha\,\,N_c\,e_q^2\;m_f^2\,\left[\ln\left(\frac{4p_c^2 q_0^2}{Q^4}\right)+2\ln 2-2\right]\nonumber
\eea
Comparing Eqns. (\ref{RS-log}) and (\ref{tempY}) we obtain $C^{\rm soft}_{2\to 2}(\xi\to\infty)=2\ln 2-2$.\\

The full numerical result for the constant $C^{\rm soft}_{2\to 2}(\xi)$ is shown in Fig. \ref{Csoft}.

\vspace*{1cm}

\par\begin{figure}[H]
\begin{center}
\includegraphics[width=10cm]{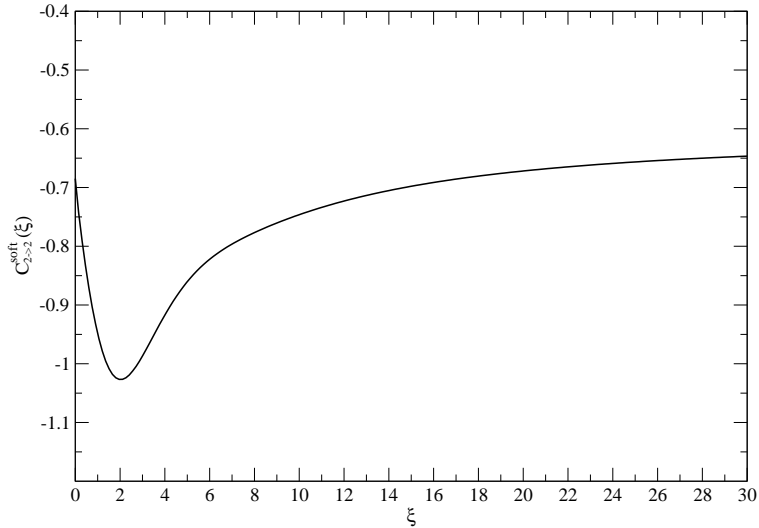}
\end{center}
\caption{The term $C^{\rm soft}_{2\to 2}(\xi)$ as defined in Eqn. (\ref{RS-log}). For $\xi \to \infty$ the result approaches $2\ln 2-2 \sim$ -0.61. For $\xi=0$ the result is approximately -0.69.}
\label{Csoft}
\end{figure}

\section{Thermal Drell-Yan Processes}
\label{DY}

\subsection{Integral Expressions}

In this section we discuss the contribution from thermal Drell-Yan processes of the form $q\bar q\to \gamma$. 
For these processes we have a hard, low virtuality photon ($\{q_0,q\}$ hard; $Q$ soft) decaying into two quarks with momenta $P$ and $R=P+Q$. By momentum conservation, at least one of these quarks must be hard. In fact, because the virtualities of both quarks are required to be soft by the kinematics, the integral is phase space suppressed unless both quarks are hard ($\{p_0,p,r_0,r\}$ hard; $\{P^2,R^2\}$ soft). The behaviour of hard, low virtuality quarks was first studied in \cite{tony}. Since we are effectively extracting a next-to-leading-order contribution from a one-loop diagram, we need to keep all next-to-leading-order terms in the asymptotic propagator. Following Ref. \cite{PAbrem}, we write:
\bea
\label{Sasym}
&&S_{\rm htl}(R) ~~ \to ~~S_{\rm asy}(R)=\frac{\Rbarsl}{R^2-m_\infty^2}, \\[2mm]
&&\Rbarsl = r_0\gamma_0-\omega_\infty(r)\,\hat r_i\gamma_i\,;~~\omega_\infty(r)=\sqrt{r^2+m_\infty^2}\approx r+m_f^2/r\,;~~m_\infty^2:=2m_f^2.\nonumber
\eea

It is normally sufficient to use bare vertices when all legs are hard. However, when calculating a next-to-leading-order contribution, one must also use corrected vertices.  One way to understand this point is to notice that the asymptotic propagator and the bare vertex do not satisfy the Ward identity. 

Combining this information, we consider the 1-loop diagram in Fig. \ref{virt}.
\par\begin{figure}[H]
\begin{center}
\includegraphics[width=4cm]{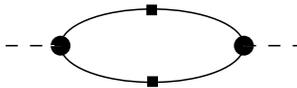}
\end{center}
\caption{The diagram from which thermal Drell-Yan processes are obtained. The quarks carry momenta $P$ and $R=P+Q$. We use the asymptotic propagators given in Eqn. (\ref{Sasym}), and the vertices given in Eqn. (\ref{gammaHtl}).}
\label{virt}
\end{figure} 
 We use corrected vertices of the form  \cite{PAbrem}:
\bea
\label{gammaHtl}
&&\Gamma_{\mu}=\big(g_{\mu\nu}+\delta \Gamma^{\rm htl}_{\mu\nu}\big)\gamma^\nu.
\eea
The explicit form of the vertex correction is not necessary. We only need to require that it satisfies the Ward identity:
\bea
\label{gammaA}
R^\mu\,\delta \Gamma^{\rm htl}_{\mu\nu} = P_\nu-\bar P_\nu\,;~~~~
P^\mu\,\delta \Gamma^{\rm htl}_{\mu\nu} = R_\nu-\bar R_\nu\,.
\eea
The integral corresponding to the diagram in Fig. \ref{virt} is constructed from Eqns. (\ref{Sasym}) and (\ref{gammaA}).

We remark that the rate for 2 $\to$ 2 processes is insensitive to asymptotic masses and vertex
corrections because the kinematics does not require both of the quarks with
momenta $R$ and $P$ to be on-shell. We have verified this by direct
calculation. 

The imaginary part of $\Pi$ is obtained by cutting both lines.  We do the
angular integral using the delta function $\delta((P+Q)^2-m_\infty^2)$ as in section \ref{ca}. After using the second delta  function to do the $p_0$ integral,
the argument of the theta function in Eqn. (\ref{theta}) becomes:
\bea
\label{thetaVirt}
g(-\omega_\infty(p)/p,p)=4 p^2 q^2-\left(Q^2-2 q_0 \omega _{\infty }(p)\right)^2
\eea
and determines the limits $p_{\rm max}$ and $p_{\rm min}$ on the $p$-integral:
\bea
\label{limApprox}
p_{\rm max}=\frac{1}{2} \left(q+ q_0 \sqrt{1-4m_\infty^2/Q^2}\right)\,,~~p_{\rm min}=\frac{1}{2} \left|q- q_0 \sqrt{1-4m_\infty^2/Q^2}\right|.
\eea
The condition $p_{\rm max}=p_{\rm min}$ immediately shows that the threshold for the production of a thermal Drell-Yan
pair is $Q^2=(2m_\infty)^2$.
\\

The integral corresponding to the diagram in Fig. \ref{virt} is given by:
\bea
\label{V-full-PA}
{\rm Im}\Pi_{\rm DY}&&= \frac{2 \alpha  e_q^2 N_c}{\pi ^2}\int dP\;\big(n_f\left(p_0\right)-n_f\left(p_0+q_0\right)\big)\\[2mm]
&&\cdot \;
d(P^2-m_\infty^2)\,d((P+Q)^2-m_\infty^2)\big[\bar P\cdot(P-\bar P)+\bar R\cdot(R-\bar R)-\bar P\cdot \bar R]\nonumber
\eea
where we have dropped terms of order $m_f^4$ and higher in the numerator, since they do not contribute to the
rate at next-to-leading-order.  
Doing the $p_0$ and angular integrals and writing the result in terms of the variable $\omega:=\omega_\infty(p)$ we obtain
\bea
\label{V-full}
{\rm Im}\Pi_{\rm DY}&& = \frac{\alpha  e_q^2 N_c}{q} \int^{\omega_{\rm max}}_{\omega_{\rm min}} d\omega\;\left(n_f\left(-\omega \right)-n_f\left(q_0-\omega\right)\right)\left(Q^2+\frac{m_f^2 \left(Q^2+2 \omega  \left(\omega -q_0\right)\right) \left(\omega ^2+\left(q_0-\omega \right)^2\right)}{\omega ^2
   \left(q_0-\omega \right)^2}\right)
\eea
where
\bea
\label{wlims}
\omega_{\rm max}=\frac{1}{2} \left(q_0+q \sqrt{1-4 m^2_{\infty }/Q^2} \right)\,;~~\omega_{\rm min}=\frac{1}{2} \left(q_0-q \sqrt{1-4 m^2_{\infty }/Q^2}\right)\,.
\eea

\subsection{Analytic Expressions}

In this section we derive some analytic approximations to Eqn. (\ref{V-full}). We are interested in two limiting cases: $q_0\gg Q\sim m_f$ and $q_0\gg Q\gg m_f$. The first limit will give the behaviour of the thermal Drell-Yan contribution close to the threshold. The second limit allows us to compare with previous calculations \cite{ruus,TT}.

\subsubsection{Threshold Expansion}

For $q_0\gg Q\sim m_f$ we have $\omega_{\rm max} \approx \omega_{\rm min} \approx q_0/2$ and thus $\omega$ is hard throughout the full range of the integration, which is consistent with our use of asymptotic propagators in Eqn. (\ref{V-full-PA}). We can expand (\ref{V-full}) in $\{q_0,\omega\}\gg \{m_f,Q\}$ and approximate 
\bea
\label{thApprox}
n_f\left(-\omega \right)-n_f\left(q_0-\omega\right)~\approx~ 1\,.
\eea
We obtain:
\bea
\label{t2}
{\rm Im}\Pi_{\rm DY}&& = \frac{\alpha  e_q^2 N_c}{q} \int^{\omega_{\rm max}}_{\omega_{\rm min}} d\omega\;\left(Q^2-\frac{2 m_f^2 \left(\omega ^2+\left(q_0-\omega \right)^2\right)}{\omega  \left(q_0-\omega \right)}\right)\,.
\eea
This result is the same as Eqn. (20) of Ref. \cite{PAbrem2} \footnote{The authors of \cite{PAbrem2} define the polarization tensor with the opposite sign. Also, there is a missing factor $1/(2q)$ in their Eqn. (20).}. This explicitly demonstrates that the contribution calculated in this section (and shown in Fig. \ref{virt}) corresponds to the first term in the LPM resummation shown schematically in Fig. \ref{oddFig}, where the vertices and quark propagators include asymptotic corrections. 

The integral in Eqn. (\ref{t2}) can be done analytically. Using Eqn. (\ref{wlims}) we obtain:
\bea
{\rm Im}\Pi_{\rm DY} =\alpha  e_q^2 N_c\left(
\sqrt{ \left(1-4 m^2_{\infty }/Q^2\right)} \left(2 m^2_{\infty }+Q^2\right) + 
2 m^2_{\infty } \ln \left[\frac{1+\sqrt{1-4m_\infty^2/Q^2}}{1-\sqrt{1-4m_\infty^2/Q^2}}\right]
   \right)\,.
\eea

In order to discuss this result, and to compare with the result of other calculations, we separate the Born term from the complete thermal Drell-Yan contribution.  
We define:
\bea
\label{bornInt4}
{\rm Im}\Pi_{\rm Born} \approx \alpha N_c e_q^2 \sqrt{Q^2 \left(Q^2-4m_\infty^2\right)}. 
\eea
and write the correction to the Born term as:
\bea
\label{cornVirt}
{\rm Im}\Pi^{(1)}_{\rm DY}:={\rm Im}\Pi_{\rm DY}-{\rm Im}\Pi_{\rm Born}.
\eea
We obtain: 
\bea
\label{V-exp}
{\rm Im}\Pi^{(1)}_{\rm DY} = 4 \alpha N_c e_q^2 m_f^2  \left(
\sqrt{1-4m_\infty^2/Q^2} ~ - ~\ln \left[\frac{1+\sqrt{1-4m_\infty^2/Q^2}}{1-\sqrt{1-4m_\infty^2/Q^2}}\right]\;
  \right).
\eea

\subsubsection{Small $m_f$ Expansion}

In the limit $q_0\gg Q\gg m_f$ we have $\omega_{\rm max} \approx q_0\,;~\omega_{\rm min}\approx Q^2/(4q_0)$, which corresponds to  $\omega_{\rm max}$ hard and $\omega_{\rm min}$ soft. 
The integral expression corresponding to the thermal Drell-Yan contribution (Eqn. (\ref{V-full})) was derived under the assumption that the leading order contribution comes from $\omega$-hard. As a consequence, we cannot get a result beyond logarithmic accuracy in the small $m_f$ limit.
Since the entire thermal Drell-Yan contribution goes to zero as $m_f\to 0$, logarithmic accuracy is sufficient. We study this limit mainly to compare with the results of other authors. 

In order to extract the log term in the small $m_f$ limit we can expand Eqn. (\ref{V-full}) in $q_0\gg\{\omega,m_f,Q\}$. Dropping the Born term we obtain:
\bea
\label{virtKln}
{\rm Im}\Pi^{(1)}_{\rm DY}\Big|_{m_f\to 0}=-2\,\alpha  \,e_q^2 \,N_c\,m_f^2\,\int^{\omega_{\rm max}}_{\omega_{\rm min}}d\omega\;\frac{1}{\omega} = -2\,\alpha  \,e_q^2 \,N_c\,m_f^2\,\ln\left(\frac{q_0^2}{Q^2}\right)\,.
\eea

\subsection{Numerical Results for the Thermal Drell-Yan Contribution}

In Fig. \ref{V-fig} we plot the results for the thermal Drell-Yan contribution to Im$\Pi$. We use  
\bea
q_0=5T,~~ T=1\,\text{GeV},~~Q=0.25\,\text{GeV}.
\eea 
In order to clarify the difference between the different calculations, we plot the correction to the Born rate only.

The solid black line is the exact numerical result for the thermal Drell-Yan contribution $\Pi_{\rm DY}^{(1)}$ obtained from Eqn. (\ref{V-full}) by subtracting the term that gives the Born contribution. 
The red dashed line is the result of patching together the analytic expressions in Eqns. (\ref{V-exp}) and (\ref{virtKln}).
The blue dot-dashed line is the result of Ref. \cite{TT}. The black dotted line (the lowest line) is the result of Ref. \cite{ruus}. In sections \ref{ruus-c} and \ref{TT-c} we discuss in detail the difference between these results and ours.
\vspace*{.8cm}
\par\begin{figure}[H]
\begin{center}
\includegraphics[width=9.7cm]{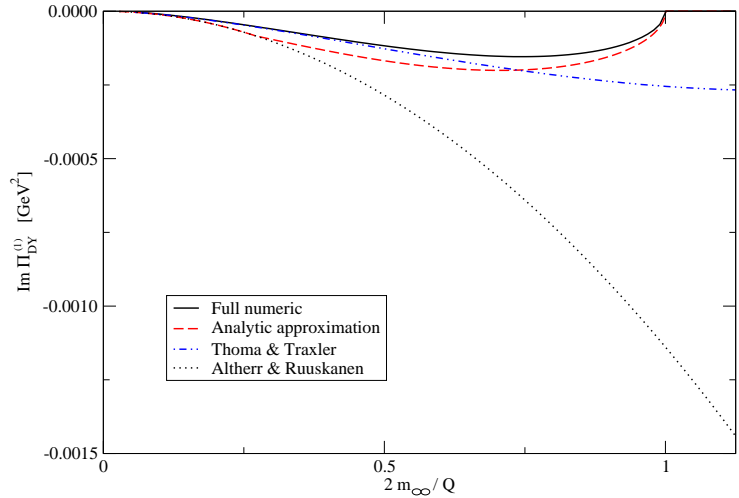}
\end{center}
\caption{The contribution $\Pi_{\rm DY}^{(1)}$. Agreement with previous results is only good for $m_f\to 0$, where $\Pi_{\rm DY}^{(1)}$ goes to zero. Our result has the correct threshold
at $Q^2=(2 m_\infty)^2$.}
\label{V-fig}
\end{figure}

 \section{Comparisons with other calculations}
 \label{COMP}

In this section we compare our results with the
results from some other papers.  This task is made more difficult by the
fact that different notation is used everywhere.  
A comparison of the definitions of $\Pi$ and the rate in the relevant papers is given in Eqn. (\ref{defnMaster}). The subscript {\em us} refers to the definitions used in this paper,  {\em AGZ} refers to the work of Aurenche at al (Refs. \cite{PAbrem,PAbrem2}); {\em AR} refers to the work of Altherr and Ruuskanen (Ref. \cite{ruus}); and {\em TT} refers to the work of Traxler and Thoma (Ref. \cite{TT}).
\bea
\label{defnMaster}
{\rm dileptons}:~~~&&{\rm rate}_{us}={\rm rate}_{AGZ}=1/2\cdot{\rm rate}_{AR}=1/2\cdot{\rm rate}_{TT},\\
{\rm Im}\Pi:~~~&&{\rm Im}\Pi_{us} = -{\rm Im}\Pi_{AGZ}= {\rm Im}\Pi_{AR}={\rm Im}\Pi_{TT} \nonumber.
\eea
Throughout the rest of this section, results from other papers are translated into our notation. 

\subsection{Comparison with the work of Altherr et al.}
\label{ruus-c}

In section \ref{ca} we use an HTL propagator for the soft exchanged quark to calculate the contribution from 2 $\to$ 2 processes, and in section \ref{DY} we use asymptotic propagators to calculate the contribution from thermal Drell-Yan processes. The authors of \cite{ruus} use expanded versions of these propagators by replacing $\frac{1}{\Psl-\Sigma} \to \frac{1}{\Psl} + \frac{1}{\Psl} \Sigma \frac{1}{\Psl}$ and consider only massless quarks. 
Writing only the log terms, their results are:
\bea
\label{ruusRes}
{\rm Im}\Pi_{DY}^{(1)} = 2 \,\alpha\,N_c\,e_q^2\, m_f^2 \ln \left(\frac{m_{\rm reg}^2}{Q^2}\right) \,;~~~~{\rm Im}\Pi_{2\to 2} = 2 \,\alpha\,N_c\,e_q^2\, m_f^2 \ln \left(\frac{q_0 T}{m_{\rm reg}^2}\right) 
\eea
where the factor $m_{\rm reg}$ is an arbitrary, unphysical regulator on the quark virtuality. The cancellation of this regulator taken to be is evidence of the KLN theorem. The results in Eqn. (\ref{ruusRes}) are equivalent to the $m_f\to 0$ limit of our results (see Eqns. (\ref{REAL}) and (\ref{virtKln})), if the regulator taken to be $m_{\rm reg}=Q^2/q_0$. 

\subsection{Comparison with the work of Thoma et al.}
\label{TT-c}

The authors of \cite{TT} intend to produce a result that is valid for $q_0\gg T\gg \{m_f, Q\}$.  
They calculate the thermal Drell-Yan contribution from a 1-loop graph, with one hard bare line, and one HTL propagator which is taken to be asymptotically hard. The diagram is multiplied by a factor of two, to account for the fact that either propagator could be the hard bare one. There are several problems with this procedure. 

Failure to use the asymptotic propagator on both lines gives the wrong symmetry factor. As a result, 
the Born rate (Eqn. (10) in Ref. \cite{TT}) is too big by a factor of two. 
In the result for the correction to the Born rate, the factor of 2 was inadvertently dropped. 

In addition, using one hard bare propagator gives the wrong threshold. The limits on the $p$-integral are obtained using a theta function like the one we have in Eqn. (\ref{thetaVirt}). Using our notation, the argument of their theta function is:
\bea
\label{thetaMT}
-\left(p^2-2 q p-Q^2-\omega_+(p)^2+2 q_0 \omega_+(p)\right) \left(p^2+2 q p-Q^2-\omega_+(p)^2+2 q_0 \omega_+(p)\right)
\eea 
Analytic approximations are obtained using the asymptotic dispersion relation $\omega_+(p)=p+m_f^2/p$ which gives:
\bea
p_{\rm max} \approx q_0\,;~~p_{\rm min} \approx \frac{2m_f^2q_0}{Q^2}.
\eea
From these expressions, one can see immediately that the threshold obtained from $p_{\rm max}=p_{\rm min}$ occurs at  $Q^2=2m_f^2=m_\infty^2$.
One can also find the threshold numerically using the full HTL dispersion replation in (\ref{thetaMT}). The result is virtually unchanged. This threshold corresponds to a decay into one quark with mass $m_\infty$ and one massless quark, or the annihilation of a massive and massless quark pair. The correct result is $Q^2=(2m_\infty)^2$, which is what we get by including the asymptotic mass on the hard line.

Finally, by neglecting vertex corrections the authors have missed some leading order contributions.

\subsection{Discussion of the results of Aurenche et al.}
\label{PA-c}

The full virtual photon spectrum has the following components:

\xx (1) The  2 $\to$ 2 processes as given in Eqn. (\ref{REAL}). We write this contribution schematically as $M_{2\to 2}$ (see Fig. \ref{caAmp}). 

\xx (2) The thermal Drell-Yan term and the associated off-shell annihilation and
bremsstrahlung processes, including multiple scattering contributions and interference terms, all of which are 
contained in the LPM resummation which was calculated in \cite{PAbrem2}. We write schematically the matrix elements corresponding to these contributions  (see Fig. \ref{oddFig}):
\bea
M^{\rm brem}_{LPM}~\sim~ \sum_{n=0}^{\infty}M_{n+1\to n+1+\gamma}\,;~~~~
M^{\rm annihil}_{LPM}~\sim~ \sum_{n=0}^{\infty}M_{n+2\to n+\gamma}\,.
\eea
The first term in the annihilation sum is $M_{2\to \gamma}$ and corresponds to the contribution from thermal Drell-Yan processes.

Refs. \cite{PAbrem2,PAbrem} dealt with bremsstrahlung and off-shell
annihilation
processes for virtual photons (see Fig. \ref{oddFig}). 
In order to compare the contributions to the overall rate from various terms, an estimate of the $2\to 2$  processes was constructed in
the following way.  The contribution from $2\to 2$ processes was taken to be the result obtained by Altherr and Ruuskanen \cite{ruus}. This result contains a $\ln(q_0 T/ Q^2)$ term and thus for small enough $Q^2$ the result of \cite{ruus}  becomes larger than the
rate calculated in \cite{joe,rolf} for real photons (which
contains a logarithmic factor $\ln(q_0 T/m_f^2)$). It is clear that this is unphysical: when $Q\to 0$ we drop below the threshold for thermal Drell-Yan processes and the rate for virtual photons should reduce to the result for real photons. 
Therefore, for small $Q^2$, the
result of \cite{ruus} was replaced by the result of \cite{joe,rolf} for real photons. As a result of this patching procedure, the
rate as a function of $Q^2$ contains a slight corner at the value of $Q^2$ where the two curves cross (see, e.g., Figs.
6 and 7 of Ref. \cite{PAbrem}). Clearly this corner is an unphysical feature.

We note that the rate calculated by Altherr and Ruuskanen
is the small $m_f$ limit of the properly calculated thermal Drell-Yan process with resummed propagators and
associated vertex corrections (Eqn. (\ref{V-full})).
Furthermore, as discussed above, 
the dressed thermal Drell-Yan process is the first (order 0) term in the
integral equation which resums the re-scattering corrections for
bremsstrahlung and off-shell annihilation, and is therefore already included
in the calculation of \cite{PAbrem2}. The thermal corrections to the Drell-Yan process have therefore been inadvertently included twice in \cite{PAbrem2,PAbrem}. 

The correct expression for the rate is the sum of the result for the LPM resummation which was calculated in \cite{PAbrem2}, and result for 2 $\to$ 2 processes as given in Eqn. (\ref{REAL}).

\section{Conclusions}
\label{conc}

We have calculated the contribution from 2 $\to$ 2 processes to the production of virtual photons (Eqn. (\ref{REAL})). Our result is valid for arbitrary values of the parameters $q_0\gg \{m_f,Q\}$. The log term is obtained analytically and agrees with the result for real photons (\cite{joe,rolf}). The constant term is obtained numerically and shown in Fig. \ref{Csoft}. In the limit $Q\to 0$ it agrees with the result for real photons. 

We have also calculated the contribution to the rate from thermal Drell-Yan processes both
analytically and numerically. The numerical result is valid for $q_0\gg T\gg
\{m_f,Q\}$, without any restrictions on the relative size on $Q$ and $m_f$. We produce two analytic expressions: one is valid for $q_0\gg Q\gg m_f$ and one is valid close to the threshold. 

The thermal Drell-Yan rate has been calculated twice previously (Refs. \cite{ruus,TT}), but neither calculation is completely correct. 
The thermal Drell-Yan contribution was also calculated in Ref. \cite{PAbrem}, as the leading order term in the LPM resummation of bremsstrahlung and off-shell annihilation processes. A slight error occured in the presentation of these results, due to a misinterpretation of the results of \cite{ruus,TT}. \\

\large

\xx {\bf Acknowledgements:}

\normalsize

\xx The authors thank Francois Gelis for useful discussions.

\end{document}